\documentstyle[12pt,axodraw]{article}
\def\PR#1#2#3{Phys. Rev. {\bf #1}, (#3) #2}
\def\PRL#1#2#3{Phys. Rev. Lett. {\bf #1}, (#3) #2}
\def\PL#1#2#3{Phys. Lett. {\bf #1}, (#3) #2}

\def\NP#1#2#3{Nucl. Phys. {\bf #1}, (#3) #2}

\def\PTP#1#2#3{Prog. Theor. Phys. {\bf #1}, (#3) #2}

\begin{document}
\title{ Zee Neutrino Mass Matrix in the Gauge Mediated
Supersymmetry Breaking Scenario}
\author{
{N. Haba$^{1}$}\thanks{E-mail address:haba@eken.phys.nagoya-u.ac.jp}
{, M. Matsuda$^2$}\thanks{E-mail address:mmatsuda@auecc.aichi-edu.ac.jp}
{ and M. Tanimoto$^3$}\thanks{E-mail
address:tanimoto@edserv.ed.ehime-u.ac.jp}
\\
\\
\\
{\small \it $^1$Faculty of Engineering, Mie University,}
{\small \it Tsu, Mie, 514-8507 Japan}\\
{\small \it $^2$Department of Physics and Astronomy,
Aichi University of Education,}\\
{\small \it Kariya, Aichi, 448-8542 Japan}\\
{\small \it $^3$Science Education Laboratory, Ehime University, }
{\small \it Matsuyama, Ehime, 790-8577 Japan}}
\date{}
\maketitle
\vspace{-10.5cm}
\begin{flushright}
hep-ph/9911511\\
\end{flushright}
\vspace{10.5cm}
\vspace{-2.5cm}
%
%
\begin{abstract}

It is well known that Zee type neutrino mass matrix can  
 provide bi-maximal neutrino mixing for three neutrinos. 
We study the reconciliation of this model with the 
 gauge mediated supersymmetry breaking scenario,  
 which naturally suppresses the large flavor changing neutral current and
CP violation in the supersymmetric standard model.
When the messenger fields have suitable $B-L$ charges,
the radiative correction naturally induces  the  Zee neutrino mass
matrix, which
provides tiny neutrino masses and  large lepton flavor mixings.
Our numerical results are consistent with the neutrino oscillation
experiments
in both three and four neutrino models.

\end{abstract}
{\sf keyword:radiative neutrino mass, large mixing, Zee model,
gauge mediated scenario}

\newpage

%
%

Recent neutrino oscillation experiments provide a strong evidence of 
tiny neutrino masses and large lepton flavor mixings\cite{solar4,Atm4,SK4}.
We know two mechanisms which can explain in a natural way the smallness 
of neutrino masses.
One is the see-saw mechanism which can induce the small neutrino
mass by integrating out heavy right-handed neutrinos\cite{see-saw}.
The second scenario is that neutrinos obtain their masses by the radiative 
corrections through which the left-handed neutrinos obtain the Majorana masses.
The latter yields small neutrino masses radiatively.
 A typical example is the so-called Zee model\cite{zee},
which does not need right-handed neutrinos\cite{zee2}.
The original Zee model is not embedded into GUT or  supersymmetry (SUSY). 
Some authors have tried to embed  it into SUSY  with $R$-parity breaking 
model\cite{Cheung}  since the right-handed slepton in SUSY has favorable quantum
number to play the role of Zee-singlet which is charged singlet
scalar under the standard model(SM). 
In these scenarios the neutrino masses strongly depend on  $R$-parity violating 
parameters in the SUSY Lagrangian and the Zee mass matrix is derived by 
the artificial adjustment of the parameters.  
\par
In this paper, we present an attractive way to embed the Zee model into 
$R$-parity conserving SUSY model.
It is well known that the gauge mediated SUSY breaking
 mechanism is one of the most reliable scenarios. 
The messenger field of SUSY-breaking can play the role of Zee-singlet which 
leads to Zee neutrino mass matrix. 
Then the neutrino masses are given in terms of the SUSY breaking parameters. 
\par
If there is no right-handed neutrino, neutrinos are unable to obtain the Dirac mass 
terms as in the SM.
In order to obtain neutrino masses without right-handed neutrinos in the SUSY model
  the following three conditions are required:  \\
\noindent
\hspace{5mm}$(i):$ $SU(2)_L$ must be broken,\\
\hspace{5mm}$(ii):$ lepton number must be broken,\\
\hspace{5mm}$(iii):$ supersymmetry must be broken.\\
\noindent
Quarks and leptons can not obtain their masses without the first condition.
The second condition is required to obtain  Majorana neutrino masses.
The lepton number conservation prevents  neutrinos from obtaining the Majorana masses
in the SM.
Recall that  neutrinos can obtain masses in the $R$-parity breaking scenario
in the SUSY theory\cite{R}.
This is due to the fact that $R$-parity is broken whenever the lepton number is broken.
The  conditions $(i)$ and $(ii)$ are also needed in the see-saw mechanism which 
includes the right-handed neutrinos.
The SUSY non-renormalization theorem requires the third condition $(iii)$
since the neutrino masses are generated by the $F$-terms in the SUSY theory.
If the SUSY is the exact symmetry, neutrinos can not obtain masses from the quantum 
corrections.
\par
Let us consider the low energy gauge mediated SUSY breaking mechanism. 
We can show that radiative corrections induce the  tiny neutrino masses and large 
lepton flavor mixing if the messenger fields have suitable $B-L$ charges, 
extra Higgs doublets, and two singlet fields have lepton number.
If  three extra singlet fields and one more pair of the messenger fields are 
added then four neutrino scenario is also realized which in particular  explains
LSND experiment\cite{LSND}.
\vspace{1cm}
\par
Taking into account the gauge mediated SUSY breaking  scenario, we investigate 
the possibility to obtain  the small neutrino masses by quantum corrections.
Gauge interaction plays the role of the  messenger of SUSY breaking in a gauge 
mediated SUSY breaking scenario, in which  the flavor changing neutral current(FCNC)  
and also $CP$ violation through the couplings of  SUSY particles are naturally suppressed. 
We introduce the singlet field $\phi$ under $SU(5)$, which is required
 to generate soft mass terms. 
This field has $F$-term as 
\begin{equation}
\label{phi}
\phi = \langle \phi \rangle +
         \langle F_{\phi} \rangle \; \theta^2 ,
\end{equation}
and SUSY breaking effects are mediated to the low energy by the couplings with
 messenger fields. 
The messenger fields ${\bf 10_M} + {\overline{\bf 10_M}}$  are introduced in 
the $SU(5)$ representation with the ordinary quantum  charge for the SM gauge 
symmetry, which component are given by 
\begin{eqnarray}
& & {\bf 10_M} = (Q_M, \overline{U_M}, \overline{E_M}), \;\;\;\;\;\;\;
{\bf \overline{10_M}} = (\overline{Q_M}, U_M, E_M) .
\end{eqnarray}
They can mediate universal soft SUSY breaking parameters through $\phi$
 by the flavor blind gauge interactions\cite{gauge-mediated}
\footnote{
The universality of soft SUSY breaking  parameters is modified,
 when there are messenger-matter mixings.
This is because the Yukawa interactions can also mediate SUSY breaking
 parameters\cite{Dine}.
Actually, soft scalar masses in our following models are shifted by the effects of
 messenger-matter mixings as discussed later.}.
The squark and slepton soft masses, and gaugino masses are given by
$(\alpha /4 \pi)( \langle F_{\phi} \rangle / \langle \phi \rangle)={\cal O}(10^2)$GeV, 
where $\alpha$ denotes gauge coupling. 
On the other hand, scalar three point soft breaking terms ($A$-terms) are 
induced by the two-loop diagrams and their magnitudes are estimated as
$(\alpha /4 \pi)^2( \langle F_{\phi} \rangle / \langle \phi \rangle)
={\cal O}(1)$GeV.
\par
The matter fields are given by
\begin{equation}
{\bf 10_f} = (Q, \overline{U}, \overline{E}), \;\;\;\;\;\;\;
{\bf \overline{5_f}} = (\overline{D}, L),
\end{equation}
which are the same as the conventional $SU(5)$ grand unified gauge theory.
The Higgs fields are given by
\begin{eqnarray}
& & \Phi = (C, H), \;\;\;\;\;\;\;
    \overline{\Phi}=( \overline{C}, \overline{H}),  \\
& & \Phi_e = (C_e, H_e), \;\;\;\;\;
\overline{\Phi_e}=
    ( \overline{C_e}, \overline{H_e}),
\end{eqnarray}
where triplets are colored Higgs $C,\overline{C}, C_e, \overline{C_e}$. 
These fields must be heavy enough to avoid rapid proton decay.
The $H$ and $\overline{H}$ are the ordinary Higgs particles,
and $H_e$ and $\overline{H_e}$ are the extra Higgs doublets.
We also introduce two gauge singlet fields $\chi$ and $\overline{\chi}$ 
with non-zero lepton number.
These fields are required in this scenario to obtain  reasonable vacuum 
expectation values of ${\cal O}(10^2)$ GeV for  $H$, $\overline{H}$, 
$H_e$, and $\overline{H_e}$.    
The extra fields (${\bf 10_M}+{\bf \overline{10_M}}$) and ($\Phi_e + 
\overline{\Phi_e}$) have the conventional gauge
quantum numbers except for unusual $B-L$ charges $Q_{B-L}$ as 
\begin{equation}
Q_{B-L} = Q_F +{2 \over 5}Y,
\end{equation}
where $Y$ is the ordinary hypercharge, and
$Q_F$-charge for the relevant fields is given in Table 1.
\begin{table}[hbtp]
\begin{center}
\begin{tabular}{|c|cccccccccc|}
\hline
  {\rm Field}   & $\Phi$ & $\overline{\Phi}$ & ${\bf \overline{5_f}}$  &
   ${\bf 10_f}$ & ${\bf 10_M}$ & ${\bf \overline{10_M}}$  &
   $\Phi_e$     & $\overline{\Phi_e}$ & $\chi$ & $\overline{\chi}$ \\
\hline
     $Q_F$  & $-{2 \over 5}$ & ${2 \over 5}$ & $-{3 \over 5}$ &
     ${1 \over 5}$ & ${6 \over 5}$ & $-{6 \over 5}$ &
     ${8 \over 5}$ & $-{8 \over 5}$ & $2$ & $-2$ \\
\hline
$Z_2$ & $+$ & $+$ & $-$ & $-$ & $+$ & $+$ & $+$ & $+$ & $+$ & $+$ \\
\hline
\end{tabular}
\caption{$Q_F$-charge in three neutrino model}
\end{center}
\end{table}
\par
In Table 1  we also introduce $Z_2$ symmetry which distinguishes matter fields 
with Higgs and messenger fields
\footnote{
$Z_2$ symmetry is the extension of the conventional $R$-parity, which distinguishes 
not only the matter field ${\bf \overline{5_f}}$ ($L$) and the Higgs field 
$\overline{\Phi}$ ($\overline{H}$), but also  the matter field ${\bf {10_f}}$ and
the messenger field ${\bf {10_M}}$.
Without  $Z_2$ symmetry  this model is similar to the  $R$-parity breaking scenario,
where neutrinos can obtain their masses through 
the mixing with neutralinos.}.
{}From the charge assignments in Table 1, we obtain 
the superpotential $W_3$ in three neutrino model as 
\begin{eqnarray}
\label{W3}
W_3 &=& {\bf 10_f} {\bf 10_f} \Phi
    +  {\bf 10_f} {\bf \overline{5_f}}\: \overline{\Phi}
    +  {\bf 10_M} {\bf \overline{5_f}}\: {\bf \overline{5_f}}
    +  {\bf 10_M} \overline{\Phi}\: \overline{\Phi_e}
    +  {\bf \overline{10_M}} \Phi \Phi_e    \nonumber \\
& & 
    +  M {\bf 10_M}{\bf \overline{10_M}}
    + \chi  \Phi \overline{\Phi_e}
    + \overline{\chi} \Phi_e \overline{\Phi}
    + \mu   \Phi \overline{\Phi}
    + \mu_e  \Phi_e \overline{\Phi_e}
    + \mu_{\chi} \chi \overline{\chi} ,
\end{eqnarray}
where the 1st and 2nd terms are usual Yukawa interactions, and the 3rd
 and 4th terms  denote the couplings of messenger field with matter and Higgs fields,
respectively. 
These  couplings are the origin to yield  Zee neutrino mass matrix radiatively. 
The 5th term is conjugate to  the 4th term. 
The 6th term corresponds to the mass term of messenger field, and 
$M$ is the order of the messenger scale induced by $\langle \phi \rangle$.
The remaining terms in $W_3$ generate Higgs scalar potential with taking 
 weak scale order for $\mu$'s.
The superpotential $W_3$ preserves $U(1)_{B-L}$ global
symmetry and $Z_2$ discrete symmetry.
\par
Notice the importance of the proton decay.
Due to $U(1)_{B-L}$ symmetry, the dominant operator to cause proton
 decay is the dimension five operator $(QQQL)$ or
$( \overline{D}\:\overline{D}\:\overline{D}\:\overline{E})$ 
mediated by the colored Higgs $C$'s or $\overline{C}$'s exchange.
We assume the triplet-doublet splittings in the Higgs sector,  where
colored Higgs fields have super-heavy masses enough to avoid the rapid proton 
decay, while Higgs doublets $H$'s and $\overline{H}$'s
have weak scale masses.
Thus the proton decay is suppressed enough in this model as in the ordinary grand
unified models. 
The gauge unification of SUSY seems to be destroyed by the existence of two extra
light Higgs doublets $H_e$ and $\overline{H_e}$.  
However, the introduction of missing partner fields could recover the
gauge coupling unification and also resolve
the triplet-doublet splitting\cite{Hisano}. 
\par
Take non-zero vacuum expectation values $\langle H_e \rangle$, $\langle
\overline{H_e}\rangle$, $\langle \chi \rangle$, and $\langle
\overline{\chi}\rangle$ at the weak scale in this model. 
Then $U(1)_{B-L}$ symmetry is spontaneously broken and massless Majoron particles 
should appear.
However, Majoron fields can almost decouple not only with quarks and leptons 
but also with gauge bosons, since Majoron fields can be almost composed of singlet fields
in this model. 
Therefore the existence of massless Majoron in this model might be  
compatible  with accelerator experiments.
\par
Now let us estimate the neutrino masses and lepton flavor
mixing angles according to Eq.(\ref{W3}).
We extract the interaction of the lepton doublet
in Yukawa coupling in the 3rd term. 
This  term is denoted by $f_{\alpha \beta}\: \overline{E_M} L_{\alpha}L_{\beta}$,
where indices correspond to the flavor: $\alpha , \beta = e, \mu, \tau$.
Since the coupling $f_{\alpha \beta}$ is antisymmetric by statics,
the diagonal elements of the neutrino mass matrix $M_{\nu}$ become  zero. 
The neutrino mass matrix is generated by Fig.1 radiatively at one loop level  
and we obtain 
\begin{equation}
M_{\nu} =
\left(
\begin{array}{ccc}
         0  & m_{e \mu}    & m_{e \tau}       \\
m_{e \mu}  &  0           & m_{\mu \tau}     \\
m_{e \tau} & m_{\mu \tau} & 0
\end{array}
\right).
\label{mnu}
\end{equation}
This is Zee type of neutrino mass matrix which leads to the stable lepton
flavor mixing matrix\cite{MNS} against
the quantum corrections\cite{HMOS}.
%
\begin{figure}[h]
\begin{center}
\mbox{
\begin{picture}(300,300)(0,0)
\ArrowLine(0,50)(40,50)
\ArrowLine(150,50)(40,50)
\ArrowLine(150,50)(260,50)
\ArrowLine(300,50)(260,50)
\DashArrowArcn(150,50)(110,90,0){3}
\DashArrowArc(150,50)(110,90,180){3}
\DashArrowLine(150,50)(150,10){3}
\DashArrowLine(150,160)(150,220){3}
\Text(150,150)[]{$A$}
\Text(70,150)[]{$\overline{H}$}
\Text(10,40)[]{$L_\alpha$}
\Text(100,40)[]{$\overline{E_\beta}$}
\Text(150,0)[]{$\langle\overline{H}\rangle$}
\Text(210,40)[]{$L_\beta$}
\Text(260,40)[]{$f_{\beta\gamma}$}
\Text(290,40)[]{$L_\gamma$}
\Text(230,150)[]{$\overline{E_M}$}
\Text(150,240)[]{$\langle\overline{H_e}\rangle$}
\end{picture}
}
\end{center}
\caption{Feynman diagram to generate neutrino masses in the three
neutrino model}
\label{loopfig1}
\end{figure}
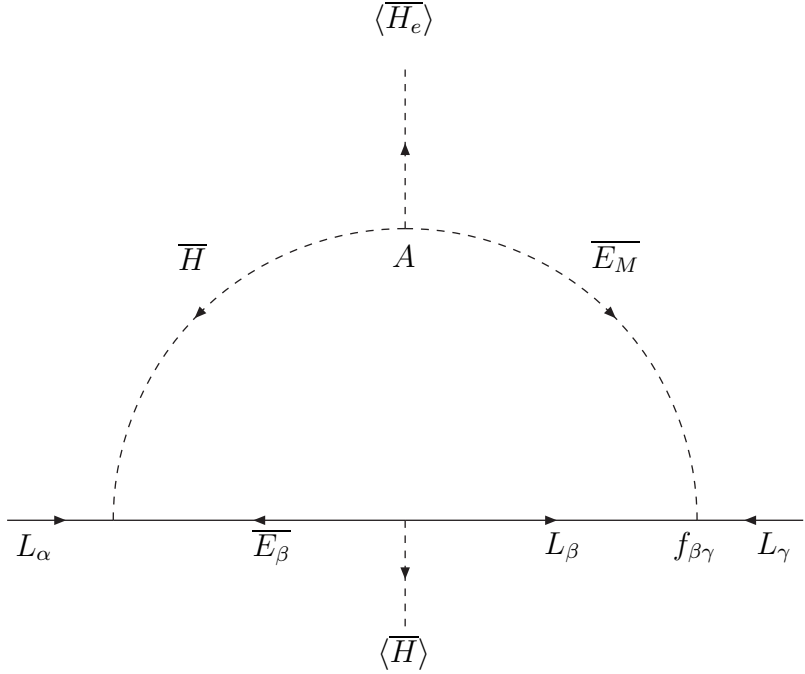
%
The neutrino masses are given as 
\begin{eqnarray}
\label{m1}
& &  m_{e \mu}= f_{e \mu}(m_{\mu}^2-m_{e}^2) A
                {\langle \overline{H_e}\rangle
                 \over \langle \overline{H}\rangle}
          F(M^2, \mu^2), \nonumber \\
& &  m_{e \tau}= f_{e \tau}(m_{\tau}^2-m_{e}^2) A
                {\langle \overline{H_e}\rangle
                 \over \langle \overline{H}\rangle}
          F(M^2, \mu^2), \\
& &  m_{\mu \tau}= f_{\mu \tau}(m_{\tau}^2-m_{\mu}^2) A
                {\langle \overline{H_e}\rangle
                 \over \langle \overline{H}\rangle}
          F(M^2, \mu^2), \nonumber
\end{eqnarray}
where
\begin{equation}
    F(M^2, \mu^2)={1 \over 16 \pi^2}{1 \over M^2- \mu^2}{\rm ln}
          {M^2 \over \mu^2} .
\end{equation}
Here $A$ is the soft mass of the scalar three point coupling
${\overline{E_M}} \; \overline{H}\; \overline{H_e}$ with 
${\cal O}(1)$GeV.
The unique mass matrix compatible with the solar and the atmospheric experiments in the
Zee model with three neutrino, requires (1,2) and (1,3) elements to be of the
same order, and (2,3) element to be negligible when compared to (1,2) and (1,3) 
elements\cite{JMST}.
When $f_{e \mu} \gg f_{e \tau} \gg f_{\mu \tau}$ and
$f_{e \mu}/ f_{e \tau}\simeq m_{\tau}^2/m_{\mu}^2$, the neutrino mass matrix 
$M_{\nu}$ in Eq.(\ref{mnu}) can induce the bi-maximal mixings,  suggesting 
the atmospheric neutrino solution and solar vacuum solution or large angle MSW solution.
The bi-maximal condition $0.02{\rm eV} < m_{e \mu} < 0.08{\rm eV}$
is realized\cite{JMST}, when $f_{e \mu}\sim 1$,
$M \sim 10^{4.5}$ GeV
\footnote{
The messenger scale of $M$ and $M_0$ in Eq.(\ref{W4}) must be larger than $10^4$ GeV.
This lower bound is required by the positivity of the scalar masses of the 
messenger quarks and leptons.
We thank Y. Mimura and Y. Nomura for this observation.} 
and $\langle \overline{H} \rangle < \langle \overline{H_e}\rangle$.
\vspace{1cm}
\par
In order to reconcile the data of LSND experiment as well as solar and atmospheric experiments,  
we need three  flavor and one sterile neutrinos at least. 
It is meaningful to construct the messenger model
 with four tiny neutrino masses by quantum corrections. 
Let us introduce another (${\bf 10_M^0}+{\bf \overline{10_M^0}}$)
messenger fields
\footnote{
In this case, the gauge couplings blow up around $10^{14}$ GeV\cite{MURA}.} 
in the $SU(5)$ gauge representation, which are denoted by
\begin{eqnarray}
& & {\bf 10_M^0} = (Q_M^0, \overline{U_M^0}, \overline{E_M^0}),
\;\;\;\;\;\;\;
{\bf \overline{10_M^0}} = (\overline{Q_M^0}, U_M^0, E_M^0).
\end{eqnarray}
Since these fields have ordinary quantum charges for the SM gauge symmetry,
they can mediate SUSY breaking through $\phi$ in Eq.(\ref{phi})
by the conventional gauge mediated scenario.
We also introduce three gauge singlet fields $S$, $N$, and $\overline{N}$
which have the lepton number. 
$S$ corresponds to the sterile neutrino.
The extra fields (${\bf 10_M}+{\bf \overline{10_M}}$), 
(${\bf 10_M^0}+{\bf \overline{10_M^0}}$), $S$, $N$, and $\overline{N}$ 
in the four neutrino case  have $Q_F$ charges as listed in Table 2 together with 
the relevant fields in the three neutrino case.
\begin{table}[hbtp]
\begin{center}
\begin{tabular}{|c|ccccccccccccccc|}
\hline
  {\rm Field}   & $\Phi$ & $\overline{\Phi}$ & ${\bf \overline{5_f}}$  &
   ${\bf 10_f}$ & ${\bf 10_M}$ & ${\bf \overline{10_M}}$  &
   $\Phi_e$     & $\overline{\Phi_e}$ & $\chi$ & $\overline{\chi}$ &
   $S$ & $N$ & $\overline{N}$ & ${\bf 10_M^0}$ & ${\bf \overline{10_M^0}}$
\\
\hline
     $Q_F$  & $-{2 \over 5}$ & ${2 \over 5}$ & $-{3 \over 5}$ &
     ${1 \over 5}$ & ${6 \over 5}$ & $-{6 \over 5}$ &
     ${8 \over 5}$ & $-{8 \over 5}$ & $2$ & $-2$ & $-1$ & $-2$ & $2$ &
     $-{4 \over 5}$ & ${4 \over 5}$ \\
\hline
$Z_2$ & $+$ & $+$ & $-$ & $-$ & $+$ & $+$ & $+$ & $+$ & $+$ & $+$ &
$-$ &
$+$ & $+$ & $+$ & $+$ \\
\hline
$Z_3$ & $1$ & $1$ & $1$ & $1$ & $1$ & $1$ & $1$ & $1$ & $1$ & $1$ &
$\omega$ & $\omega$ & $\omega^2$ & $\omega$ & $\omega^2$ \\
\hline
\end{tabular}
\caption{$Q_F$-charge in four neutrino model}
\end{center}
\end{table}
Here we also introduce $Z_3$ symmetry, which avoids the tree level mass of 
sterile neutrino $S$, for example, through the term $\chi\chi S$.
The charge assignment in Table 2 determines the superpotential $W_4$ 
in the four neutrino model as 
\begin{eqnarray}
\label{W4}
W_4 &=& W_3
    +  S {\bf 10_f} {\bf \overline{10_M^0}}
    +  N {\bf 10_M} {\bf \overline{10_M^0}}
    +  M_0 {\bf 10_M^0}{\bf \overline{10_M^0}}
    + \mu_N  N \overline{N},
\end{eqnarray}
where $W_3$ is given by Eq.(\ref{W3}). 
The 2nd and 3rd terms give the mass mixings between sterile and active neutrinos 
as shown in Fig.2. 
The 4th term is the mass term of ${\bf 10_M^0}$ and ${\bf \overline{10_M^0}}$ 
and $M_0$ is of the same order as the messenger scale. 
The last term is also the mass term of $N$ and $\overline{N}$, where 
$\mu_N$ is settled around the weak scale.  
The superpotential $W_4$ preserves $Z_2 \times Z_3$ symmetry.
In the four neutrino model we also assume that the triplet-doublet splittings
are realized in the Higgs sector, and the Higgs fields $H, \overline{H},H_e, \overline{H_e}$
and the singlet field $N$ have vacuum expectation values of order of the weak scale.
\begin{figure}[h]
\begin{center}
\mbox{
\begin{picture}(300,300)(0,0)
\ArrowLine(0,50)(40,50)
\ArrowLine(150,50)(40,50)
\ArrowLine(150,50)(260,50)
\ArrowLine(300,50)(260,50)
\DashArrowArcn(150,50)(110,90,0){3}
\DashArrowArc(150,50)(110,90,180){3}
\DashArrowLine(150,50)(150,10){3}
\DashArrowLine(150,160)(150,220){3}
\Text(150,150)[]{$A'$}
\Text(70,150)[]{$E_M^0$}
\Text(10,40)[]{$S$}
\Text(40,40)[]{$k_\alpha$}
\Text(80,40)[]{$\overline{E_\alpha}$}
\Text(150,0)[]{$\langle\overline{H}\rangle$}
\Text(200,40)[]{$L_\alpha$}
\Text(260,40)[]{$f_{\alpha\beta}$}
\Text(290,40)[]{$L_\beta$}
\Text(220,150)[]{$\overline{E_M}$}
\Text(150,240)[]{$\langle N\rangle$}
\end{picture}
}
\end{center}
\caption{Feynman diagram to generate neutrino masses in the four
neutrino model}
\label{loopfig2}
\end{figure}
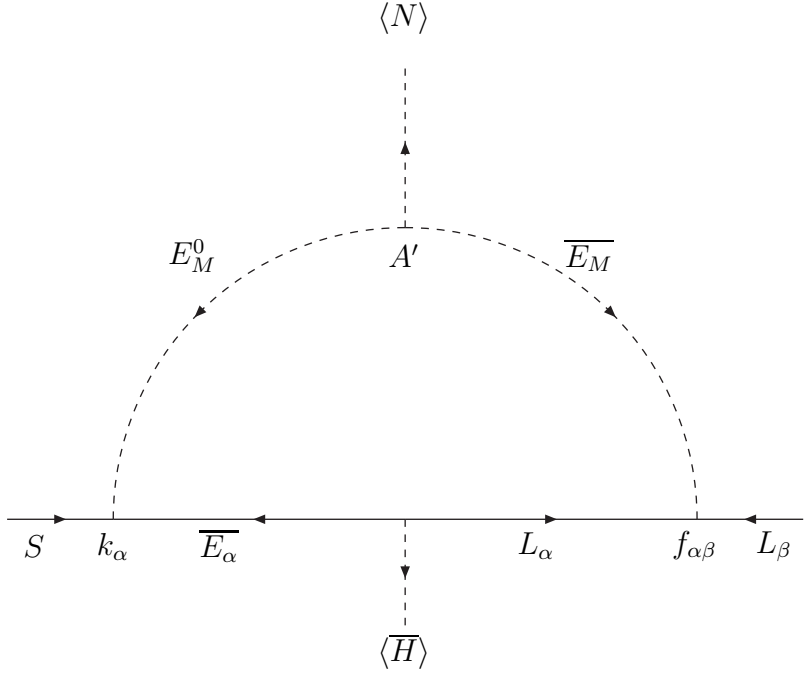
\par
Next we estimate the neutrino masses and lepton flavor mixing angles in the 
four neutrino model.
We extract the interaction of the lepton due to  the 2nd term in Eq.(\ref{W4}), 
which is denoted as $k_{\alpha}S \overline{E_{\alpha}} E_M^0$.
The $4 \times 4$ neutrino mass matrix is given by
\begin{equation}
M_{\nu} =
\left(
\begin{array}{cccc}
         0  & m_{e \mu}    & m_{e \tau}   & m_{e s}    \\
m_{e \mu}  &  0           & m_{\mu \tau} & m_{\mu s}  \\
m_{e \tau} & m_{\mu \tau} & 0            & m_{\tau s} \\
m_{e s}    & m_{\mu s}    & m_{\tau s}   & 0
\end{array}
\right).
\label{m_nu}
\end{equation}
The mass terms including the sterile neutrino field $S$\cite{sterile}
are obtained from the diagram of Fig.2 as
\begin{eqnarray}
& &  m_{e s}= (f_{e \tau}k_{\tau}m_{\tau}+
               f_{e \mu} k_{\mu} m_{\mu} ) A'
              \langle N \rangle F(M^2, M_0^2), \nonumber \\
& &  m_{\mu s}= (f_{\mu \tau}k_{\tau}m_{\tau}+
               f_{\mu e} k_{e} m_{e} ) A'
              \langle N \rangle F(M^2, M_0^2), \\
& &  m_{\tau s}= (f_{\tau \mu}k_{\mu}m_{\mu}+
               f_{\tau e} k_{e} m_{e} ) A'
              \langle N \rangle F(M^2, M_0^2), \nonumber
\end{eqnarray}
where $A'$ is the soft mass of the scalar three point coupling
$N \overline{E_M} E_M^0$, which is of order 1 GeV.
The same type of mass matrix as in Eq.(\ref{m_nu}) was
already analyzed in details in Ref.\cite{Roy}.
\par
When $f_{e \mu} \ll f_{e \tau} < f_{\mu \tau}$ and 
$k_{\tau} \ll k_{\mu} \leq k_e$, this model can explain the LSND as well as 
atmospheric and the solar neutrino experiments, where the small mixing 
angle MSW solution is preferred. 
{}For example, when $f_{\mu \tau}\simeq 1$, $f_{e \tau}\sim 0.1$, 
$f_{e \mu}\simeq 10^{-3}$,
$k_e = k_{\mu} \simeq 1$ and $k_{\tau}\simeq 10^{-3}$  the elements might be
$m_{\mu \tau}= 0.5$ eV, $m_{e \tau}= 0.05$ eV,
$m_{e \mu}= 10^{-5}$ eV, $m_{\tau s}= 0.15$ eV,
$m_{\mu s}= 0.0036$ eV, and $m_{e s}= 0.00025$ eV,
which in turn, induce suitable neutrino mass squared differences
$\delta m^2_{\rm sol}= 4 \times 10^{-6}$ eV$^2$,
$\delta m^2_{\rm atm}= 2 \times 10^{-3}$ eV$^2$,
$\delta m^2_{\rm LSND}= 0.3$ eV$^2$,
and mixing angles
$\sin^2 2 \theta_{\rm sol} = 1 \times 10^{-3}$,
$\sin^2 2 \theta_{\rm atm} = 0.9$,
$\sin^2 2 \theta_{\rm LSND} = 0.03$\cite{Roy}.
\vspace{1cm}
\par
Let us summarize the results.
The Zee neutrino mass matrix is naturally realized in the three neutrino 
scenario in the frame of SUSY theory provided that the messenger fields 
have suitable $B-L$ charges, and extra Higgs doublets 
($H_e$ and $\overline{H_e}$) and two singlet fields 
($\chi$ and $\overline{\chi}$) have lepton numbers.
If three extra singlet fields and one more pair of the messenger fields are 
added then the four neutrino scenario is realized.
This mass matrix is consistent with solar, atmospheric and LSND
experiments when the parameters are chosen appropriately.
\par
We shall comment now on the messenger-matter mixings. 
The most advantage of gauge mediated SUSY breaking is the natural
derivation of universal soft masses which are flavor blind as well
known\cite{Dine}.
However, the interaction $f_{\alpha\beta}10_M{\overline 5_f}_\alpha{\overline 5_f}_\beta$ 
induces the mixing between messenger and matter field and the universality 
of soft mass might be broken by this term.
Especially the soft mass terms between the first and the second generation are
strongly degenerated  by the constraints of the experiments on $K-{\overline K}$ mixing and
$\mu\rightarrow e\gamma$ decay.
According to Ref.\cite{Dine}, the soft mass for ${\tilde D}$ and 
${\tilde L}$ can receive the correction as ${\cal O}(F_\phi^2/M^4)$. 
In the present model for three neutrino case we have the relation
$f_{e\mu}\gg f_{e\tau} \gg f_{\mu\tau}$.
Due to this relation  the corrections for first and  second sleptons are 
common and the model is safe from the above constraints.
For four neutrino models there exists new term  $S10_f{\overline
{10_M^0}}$, which gives the shifts of soft masses for ${\tilde{\overline E}}$, ${\tilde Q}$ and
${\tilde{\overline U}}$. 
Though the constraints from the experiments are rather severe under 
the relation $f_{e\mu} \ll f_{e\tau} < f_{\mu\tau}$, it is expected that 
the order of mass shift can be phenomenologically acceptable 
since $k_e = k_{\mu} \simeq 1$. 
This provides the suitable neutrino mass matrix as we argue in this paper.

\section*{Acknowledgment}

One of the authors(N.H) would like to thank S. Raby, K. Tobe, M. Schmarz,
M. Strasster, Y. Nomura, and Y.Mimura for useful discussions at the early 
stage of this work.
We also give our thanks for kind hospitality at Summer Institute for 
Neutrino Physics at Fuji-yoshida in August, 1999. This work was mainly done through 
this Summer Institute.   
Authors thank T.Turova for careful reading of the manuscript.
%
%


\begin{thebibliography}{99}
\bibitem{solar4}
Homestake Collaboration, B.T.~Cleveland {\it et al.},
  Nucl.~Phys.(Proc.~Suppl.) {\bf B38}, (1995) 47;
Kamiokande Collaboration, Y.~Suzuki,
  Nucl.~Phys. (Proc.~Suppl.) {\bf B38}, (1995) 54;
GALLEX Collaboration,  P.~Anselmann {\it et al.},
\PL{B357}{237}{1995};
SAGE Collaboration, J.N.~Abdurashitov {\it et al.},
\PL{B328}{234}{1994};
Super-Kamiokande Collaboration,
\PRL{82}{1810}{1999}; \PRL{82}{2430}{1999}; hep-ex/9903034.

\bibitem{Atm4}
Kamiokande Collaboration, K.S.~Hirata {\it et al.}, \PL{B205}{416}{1988};
{\it
  ibid.} {\bf B280}, 146 (1992); Y.~Fukuda {\it et al.},
  \PL{B335}{237}{1994};
IMB Collaboration, D.~Casper {\it et al.},
  \PRL{66}{2561}{1991}; R.~Becker-Szendy {\it et al.}, \PR{D46}{3720}{1992};
SOUDAN2 Collaboration, T.~Kafka, Nucl.~Phys. (Proc.~Suppl.) {\bf B35}, 427
  (1994); M.C.~Goodman, {\it ibid.} {\bf 38}, (1995) 337; W.W.M.~Allison
{\it
  et al.}, \PL{B391}{491}{1997}; \PL{B449}{137}{1999}.

\bibitem{SK4}
Y.~Totsuka, invited talk at the 18th International Symposium on
Lepton-Photon
  Interaction, July 28 - August 1, 1997 Hamburg;
Super-Kamiokande Collaboration,
\PRL{81}{1562}{1998}; \PRL{82}{2644}{1999}; hep-ex/9903047.


\bibitem{see-saw}
T. Yanagida, in {\it Proceedings of the Workshop on the Unified Theories and
Baryon Number in the Universe}, Tsukuba, Japan, 1979, edited by O. Sawada
and A. Sugamoto, KEK Report No. 79-18, Tsukuba, 1979, p.95;
M. Gell-Mann, P. Ramond and R. Slansky, in {\it Supergravity},
Proceedings of the Workshop, Stony Brook, New York, 1979,
edited by P. van Nieuwenhuizen and D. Freedmann, North-Holland,
Amsterdam, 1979, p.315.



\bibitem{zee}
A. Zee, \PL{93}{389}{1980}; \PL{161}{141}{1985}.


\bibitem{zee2}
L.~Wolfenstein, \NP{B175}{92}{1980};
S. ~T. ~Petcov, \PL{B115}{401}{1982};
J. ~Liu, \PL{B216}{367}{1989};
W.~Grimus and H.~Neufeld, \PL{B237}{521}{1990};
B.~K.~Pal, \PR{D44}{2261}{1991};
W.~Grimus and G.~Nardulli, \PL{B271}{161}{1991};
A.~Yu.~Smirnov and Z.~Tao, \NP{B426}{415}{1994};
A.~Yu.~Smirnov and M. ~Tanimoto, \PR{D55}{1665}{1997};
C.~Jarlskog, M.~Matsuda, S.~Skadhauge, M.~Tanimoto, \PL{B449}{240}{1999};
P.~Frampton, S.~L.~Glashow,\PL{B461}{95}{1999}.

\bibitem{Cheung}
K. Cheung and O. C. W. Kong, hep-ph/9912238.

\bibitem{gauge-mediated}
M.~Dine and A.~E.~Nelson, \PR{D48}{1277}{1993};
M.~Dine, A.~E.~Nelson, and Y.~Shirman, \PR{D51}{1362}{1995};
M.~Dine, A.~E.~Nelson, Y. ~Nir, and Y.~Shirman, \PR{D53}{2658}{1996}.

\bibitem{Dine}
M.~Dine, Y.~Nir, and Y.~Shirman, \PR{D55}{1501}{1997}.


\bibitem{R}
L.J. Hall and M. Suzuki, Nucl. Phys. {\bf B231} (1984) 419;
I. Lee, Nucl. Phys. {\bf B246} (1984) 120;
V. Barger, G.F. Giudice and T. Han, Phys. Rev. {\bf D40} (1989) 2987;
K. Enqvist, A. Masiero and A. Riotto, Nucl. Phys. {\bf B373} (1992) 95;
J.C. Romao and J.W.F. Valle, Nucl. Phys. {\bf B381} (1992) 87;
F. Vissani and A.Yu. Smirnov, Nucl. Phys. {\bf B460} (1996) 37;
R.~Hempfling,\NP{B478}{3}{1996};
K.S. Babu and R.N. Mohapatra, Phys. Rev. Lett. {\bf B384} (1996) 123;
B. de Carlos and P.L. White, Phys. Rev. {\bf D54} (1996) 3424;
A. Akeroyd, M.A. Diaz, J. Ferrandis, M.A. Garcia-Jareno and J.W.F. Valle,
Nucl. Phys. {\bf B529} (1998) 3;
V. Bednvyakov, A. Faessler and S. Kovalenko,
Phys. Lett. {\bf B442} (1998) 203;
M. Mukhopadhyaya, S. Roy and F. Vissani,
Phys. Lett. {\bf 443B}, (1998) 191;
M.~Hirsch, H.~V. ~Klapdor-Kleingrothaus,  S.~G. ~Kovalenko,
\PR{D57}{1947}{1998};
E.J. Chun, S.K. Kang, C.W. Kim and U.W. Lee,
Nucl. Phys. {\bf B544} (1999) 89;
C.~Liu and H.S.~Song, \NP{B545}{183}{1999}.
R. Adhikari and G. Omanovic, Phys. Rev. {\bf D59} (1999) 073003;
A. Raychaudhuri, Phys. Rev. {\bf D59} (1999) 091701;
K. Choi, E.J. Chun and K. Hwang, Phys. Rev. {\bf D60} (1999) 031301;
E.J. Chun and J.S. Lee, Phys. Rev. {\bf D60} (1999) 075006;
G.~Bhattacharyya, H.~V.~Klapdor-Kleingrothaus, and H. ~Pas,
\PL{463B}{77}{1999};
B. Mukhopadhyaya, hep-ph/9907275;
A. Abada and M. Losada, hep-ph/9908352;
S. Rakshift, G. Bhattacharyya and E. J. Chun and S. K. Kang, hep-ph/9909429;
F. Takayama and M. Yamaguchi, hep-ph/9910320.


\bibitem{LSND}
LSND Collaboration, C. Athanassopoulos et al., Phys. Rev. Lett.
{\bf 81} (1998) 1774;
Phys. Rev. {\bf C58} (1998) 2489.

\bibitem{Hisano}
J. Hisano, T. Moroi, T. Tobe and T. Yanagida, 
\PL{B342}{138}{1995};

\bibitem{MNS}
Z.~Maki, M.~Nakagawa and S.~Sakata, \PTP{28}{870}{1962}.


\bibitem{HMOS}
N.~Haba and N.~Okamura, hep-ph/9906481;
N.~Haba, M.~Matsui, N.~Okamura, and M.~Sugiura, hep-ph/9908429.


\bibitem{JMST}
C.~Jarlskog et. al. and P.~Frampton et. al. in Ref.\cite{zee2}.


\bibitem{MURA}
C.~D.~Carone and H. ~Murayama, \PR{D53}{1658}{1996}.

\bibitem{sterile}
V. Barger, P. Langacker, J. Leveille, and S. Pakvasa, Phys. Rev.
  Lett.{\bf 45} (1980) 692;
J. T. Peltoniemi, D. Tommasini and J. W. F. Valle, Phys. Lett.  {\bf B298}
(1993) 383;
J. T. Peltoniemi and J. W. F. Valle, Nucl. Phys. {\bf B406} (1993) 409;
D. O. Caldwell and R. N. Mohapatra,  Phys. Rev. {\bf D48} (1993) 3259;
E. J. Chun et al.,  Phys. Lett. {\bf B357} (1995) 608;
E. Ma and P. Roy,  Phys. Rev.  {\bf D52} (1995) R4780;
J. J. Gomez-Cadenas and M. C. Gonzalez-Garcia, Z. Phys. {\bf C71} (1996)
443;
A. Yu. Smirnov and M. Tanimoto in Ref.\cite{zee2};
S. Goswami,  Phys. Rev. {\bf D55} (1997) 2931;
K. Benakli and A. Yu. Smirnov,  Phys. Rev. Lett. {\bf 79} (1997) 4314;
N. Okada and O. Yasuda,  Int. J. Mod. Phys.  {\bf A12} (1997) 3669;
S. C. Gibbons et al.,  Phys.  Lett. {\bf B430} (1998) 296;
S. Mohanty, D. P. Roy and U. Sarkar, Phys. Lett. {\bf B445} (1998) 185;
Q. Y. Liu and A. Yu. Smirnov, Nucl. Phys. {\bf B524} (1998) 505;
N. Gaur  et al., Phys. Rev. {\bf D58}  (1998) 071301;
E. J. Chun, C. W. Kim and U. W. Lee,  Phys. Rev. {\bf D58} (1998) 093003;
S. M. Bilenky, C. Giunti, W. Grimus and T. Schwetz, Astropart. Phys. {\bf 11} (1999) 413, 
Phys. Rev.{\bf D60} (1999) 073007;
S. M. Bilenky and C. Giunti, hep-ph/9905246.


\bibitem{Roy}
N.~Gaur, A.~Ghosal, E.~Ma, and P.~Roy, \PR{D58}{R7131}{1998}.

\end{thebibliography}
\end{document}